\documentclass[twocolumn]{aastex63}

\usepackage{amsmath}
 \usepackage{graphicx}
 
\shorttitle{Constraining the Milky Way Mass with Its Corona}
\shortauthors{GUO ET AL.}

\defcitealias{henley13}{HS13}
\defcitealias{bland16}{BG16}

\begin{document}
\bibliographystyle{aasjournal}

\title {Constraining the Milky Way Mass with Its Hot Gaseous Halo}

\author{Fulai Guo}
\affiliation{Key Laboratory for Research in Galaxies and Cosmology, Shanghai Astronomical Observatory, Chinese Academy of Science, 80 Nandan Road, Shanghai 200030, China; fulai@shao.ac.cn}
\affiliation{School of Astronomy and Space Science, University of Chinese Academy of Sciences, 19A Yuquan Road, 100049, Beijing, China}

\author{Ruiyu Zhang}
\affiliation{Key Laboratory for Research in Galaxies and Cosmology, Shanghai Astronomical Observatory, Chinese Academy of Science, 80 Nandan Road, Shanghai 200030, China; fulai@shao.ac.cn}
\affiliation{School of Astronomy and Space Science, University of Chinese Academy of Sciences, 19A Yuquan Road, 100049, Beijing, China}

\author{Xiang-Er Fang}
\affiliation{Key Laboratory for Research in Galaxies and Cosmology, Shanghai Astronomical Observatory, Chinese Academy of Science, 80 Nandan Road, Shanghai 200030, China; fulai@shao.ac.cn}

\begin{abstract}
We propose a novel method to constrain the Milky Way (MW) mass $M_{\rm vir}$ with its corona temperature observations.  For a given corona density profile, one can derive its temperature distribution assuming a generalized equilibrium model with non-thermal pressure support. While the derived temperature profile decreases substantially with radius, the X-ray-emission-weighted average temperature, which depends most sensitively on $M_{\rm vir}$, is quite uniform toward different sight lines, consistent with X-ray observations. For an Navarro-Frenk-White (NFW) total matter distribution, the corona density profile should be cored, and we constrain $M_{\rm vir}=(1.19$ - $2.95) \times 10^{12} M_{\sun}$. For a total matter distribution contributed by an NFW dark matter profile and central baryons, the corona density profile should be cuspy and $M_{\rm vir,dm}=(1.34$ - $5.44) \times 10^{12} M_{\sun}$. Non-thermal pressure support leads to even higher values of $M_{\rm vir}$, while a lower MW mass may be possible if the corona is accelerating outward. This method is independent of the total corona mass, its metallicity, and temperature at very large radii.  
\end{abstract}


\section{Introduction}
\label{section:intro}

During cosmic structure formation, dark matter (DM) and baryonic particles fall into existing gravitational potential wells. Within the virial radius ($r_{\rm vir}$) of a gravitating halo, it is often assumed that particles are virialized and lose memory of initial conditions, reaching a dynamical equilibrium. Under this approximation, the halo matter distribution can be measured through the Jeans equation for collisionless particles \citep{binney08}, such as stars, globular clusters and satellite galaxies, and through the hydrostatic equilibrium (HSE) equation for collisional particles such as hot gas \citep{allen11,kravtsov12}. The former method has been used extensively, including to measure the MW mass $M_{\rm vir}$ (\citealt{bland16}, hereafter BG16; \citealt{wang20}), while the latter has been used to measure the mass profiles of massive elliptical galaxies and galaxy clusters \citep{allen11,kravtsov12}.
 
X-ray observations of galaxy clusters often measure the radial temperature and density profiles of the hot halo gas up to about $0.5r_{\rm vir}$ \citep{vikhlinin06} and recently even up to $r_{\rm vir}$ in some systems \citep{ghirardini19}. Assuming HSE and spherical symmetry, gravitating masses $M(r)$ within a given radius $r$ can then be determined from thermal pressure gradients, and the thus measured cluster masses have been used extensively to constrain cosmological parameters \citep{allen11,kravtsov12}. Mounting multi-wavelength observations indicate that there exists a hot corona surrounding our MW, possibly extending to $r_{\rm vir}$ and accounting for a substantial fraction of its missing baryons (\citealt{fang13}; \citetalias{bland16}; \citealt{bregman18}). However, the MW corona properties have not yet been used to measure $M_{\rm vir}$, partly due to the low corona density and surface brightness. Furthermore, our special location near the center of the MW halo makes it very difficult, if possible, to measure the radial density and temperature distributions of the corona gas. 

The MW mass $M_{\rm vir}$ is a fundamental quantity in astronomy. While it has been measured extensively with collisionless objects, it is still uncertain to more than a factor of two due to limited number or spatial coverage of kinematic tracers (\citetalias{bland16}; \citealt{wang20}). The accurate determination of $M_{\rm vir}$ is important, as it affects if a large fraction of baryons are missing in the MW (\citealt{fang13}; \citetalias{bland16}; \citealt{bregman18}) and if there is a serious ``too-big-to-fail" problem for the MW satellites, which may challenge the cold DM theory \citep{boylan12}. Here we propose a novel method to constrain $M_{\rm vir}$ based on the properties of the collisional hot gas in the MW corona, and demonstrate that the corona temperature measurements from X-ray observations can be used to put constraints on $M_{\rm vir}$.

The virial theorem provides a crude $M_{\rm vir}-$dependent estimate of the corona temperature at $r_{\rm vir}$: $T_{\rm vir}\sim 5\times 10^{5}(M_{\rm vir}/10^{12}M_{\sun})^{2/3}$ K. At $r<r_{\rm vir}$, $T$ further rises due to adiabatic compression and heatings by turbulence, shocks, stellar feedback and active galactic nucleus (AGN) feedback. However, if the gas temperature is too high, the MW gravity could not hold the gas for a given density distribution, leading to the corona expansion and a decrease in temperature. This argument is manifested in a generalized HSE equation $dP/dr=-(1-f_{\rm nt})G\rho M(r)/r^{2}$, which may be rewritten as
\begin{eqnarray}
-\frac{d \text{~ln}T}{d \text{~ln}r}-\frac{d \text{~ln}\rho}{d \text{~ln}r}=(1-f_{\rm nt}) \frac{\mu m_{\mu}}{k_{B} T}\frac{GM(r)}{r} \text{~.}  \label{hydro1}
\end{eqnarray}
\noindent
Here $\rho$, $T$, and $P$ are the gas density, temperature, and thermal pressure respectively. $k_{B}$ is Boltzmann's constant, $G$ is the gravitational constant, $m_{\mu}$ is the atomic mass unit, and $\mu = 0.61$ is the mean molecular weight. $f_{\rm nt}$ ($0\leq f_{\rm nt} \leq 1$) is a potentially radius-dependent parameter representing the impact of non-thermal pressure support. Following any disturbance on scale $L$, the corona will return back to equilibrium quickly after a sound crossing time $t_{\rm s}\equiv L/c_{\rm s}\sim 4.6 (L/1 \text{~kpc})(T/2\times 10^{6}\text{~K})^{-0.5}$ Myr. 
 
 \begin{figure*}
   \begin{center}
\includegraphics[width=0.3\textwidth]{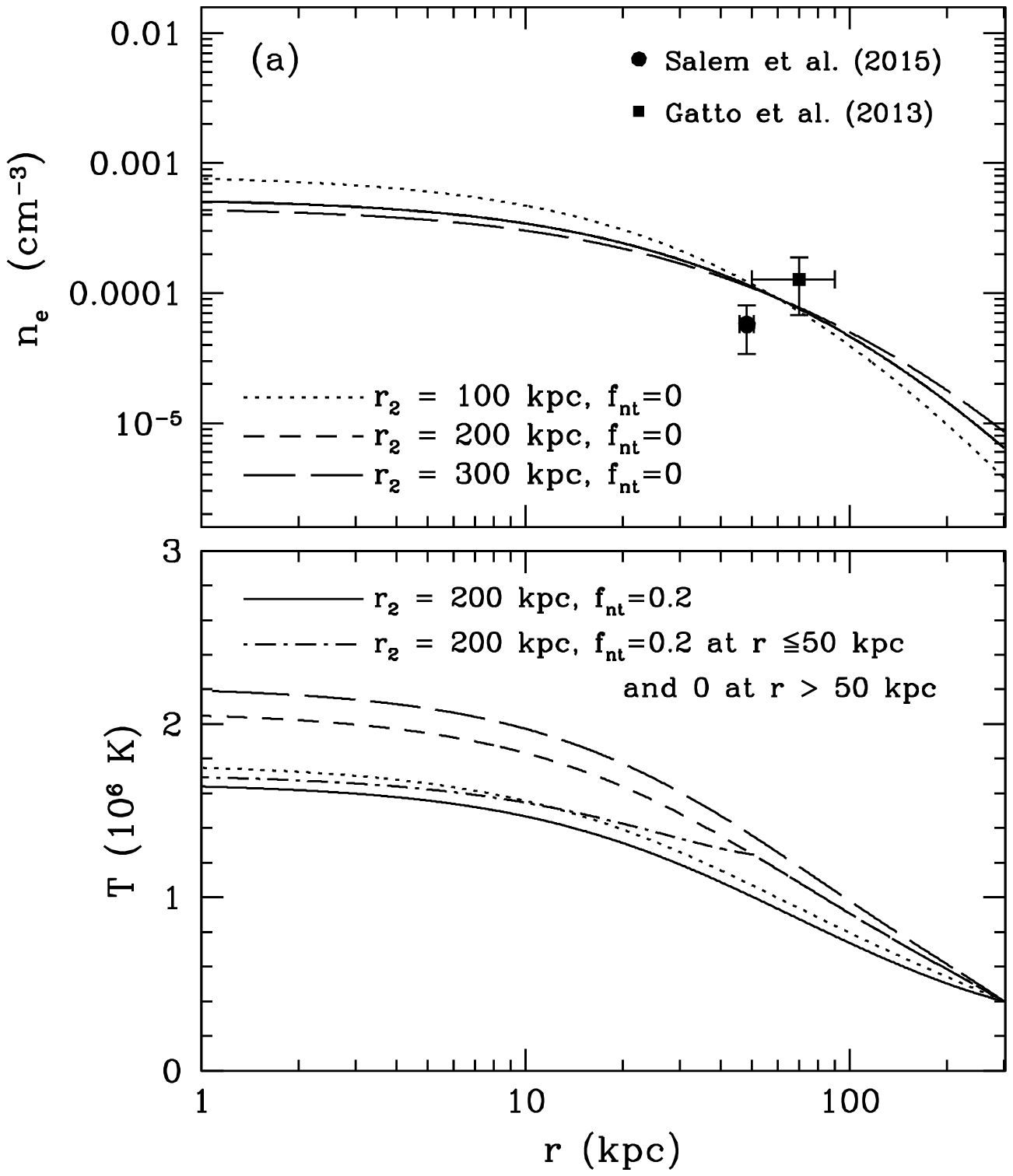} ~~~~
\includegraphics[width=0.3\textwidth]{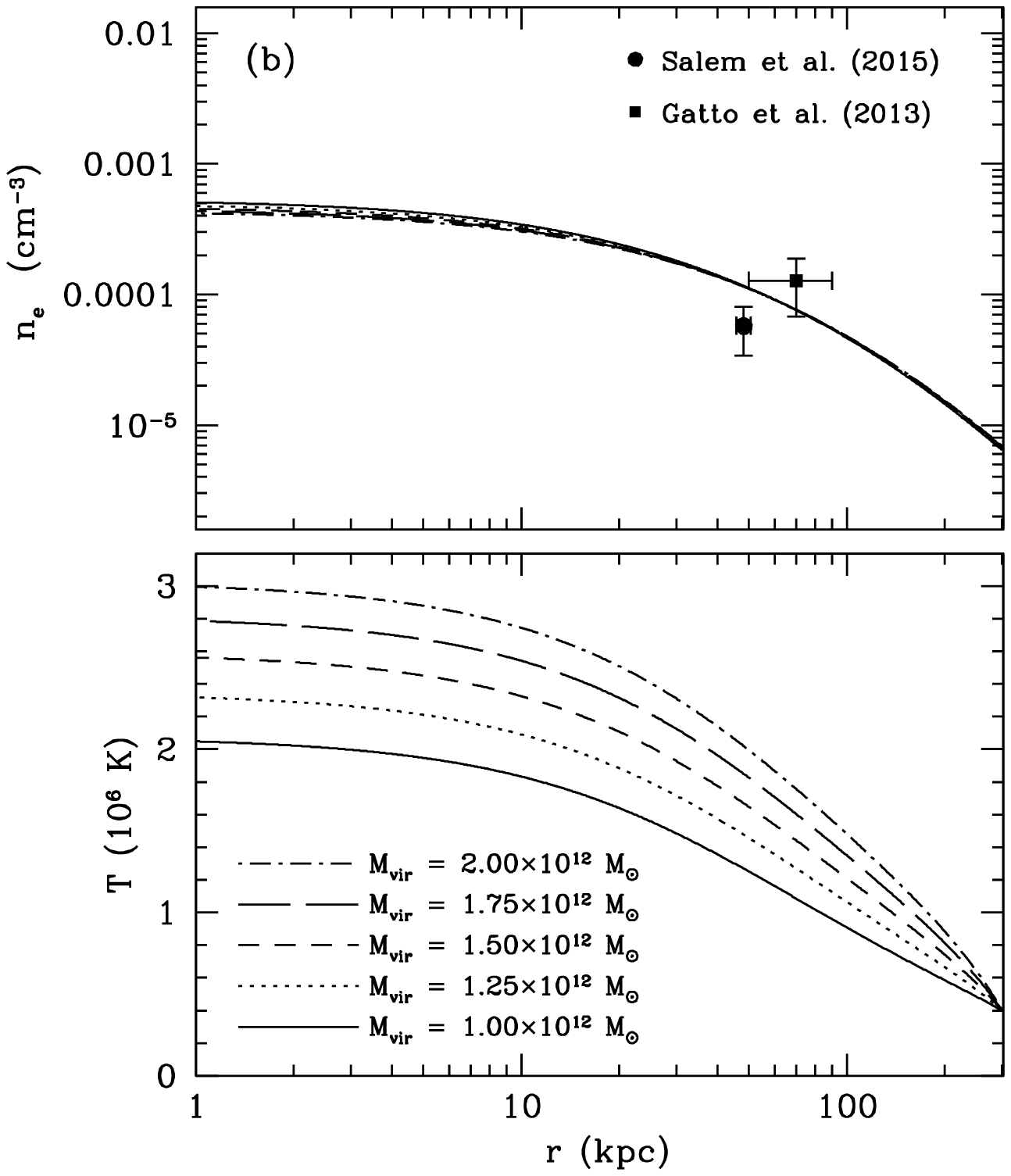} ~~~~
\includegraphics[width=0.3\textwidth]{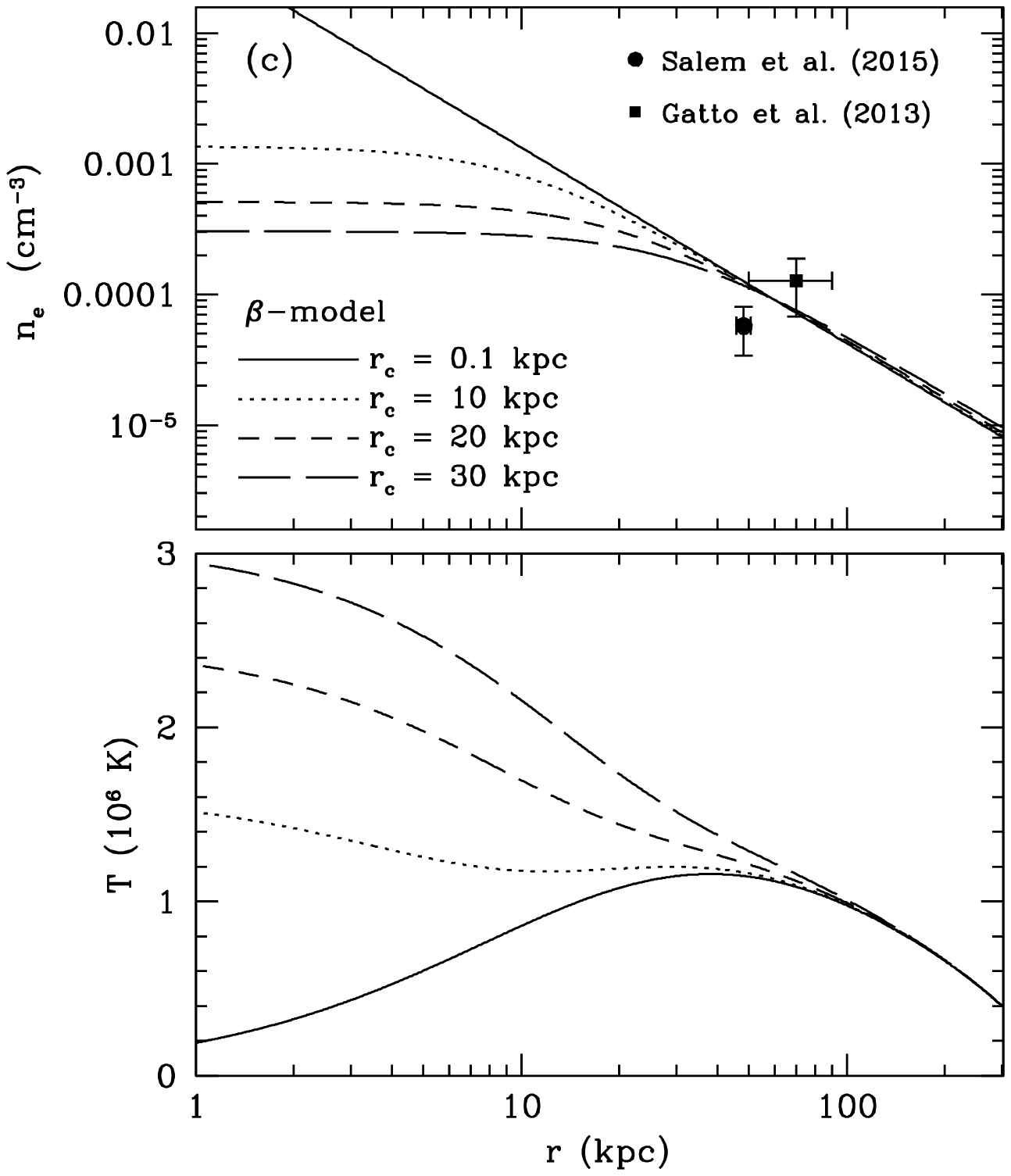} 
  \end{center}
  \vspace{-0.5cm}
\caption{Radial distributions of electron number density (top) and temperature (bottom) in (a) five default models with $M_{\rm vir}=10^{12} M_{\sun}$ and varying values of $r_{2}$ and $f_{\rm nt}$, (b) five hydrostatic models with $f_{\rm nt}=0$, $r_{2}=200$ kpc and varying values of $M_{\rm vir}$, and (c) four $\beta$ models with $M_{\rm vir}=10^{12} M_{\sun}$, $f_{\rm nt}=0$, $\beta=0.5$, and varying values of $r_{\rm c}$. Note that the value of $f_{\rm nt}$ does not affect our corona density model. The solid square and circle data points correspond to recent density estimates \citep{gatto13,salem15}.}
 \label{plot1}
 \end{figure*} 
 
\section{Method}
\label{section:method}

To constrain $M_{\rm vir}$ with the corona temperature, one needs to adopt a MW total matter distribution and a corona density distribution. The corona temperature distribution can then be solved from Equation (\ref{hydro1}) starting from an outer boundary $r_{\rm out}=300$ kpc. The gas temperature at $r_{\rm out}$ is assumed to be $T_{\rm out}=4\times 10^{5}$ K, which has little impact on the derived temperature profile in the inner region $r\lesssim 50$ kpc. In our default models, we adopt the NFW profile \citep{navarro1996,navarro97} for the MW total matter distribution, which contains two parameters: $M_{\rm vir}$ and the concentration $c$. Throughout this paper, $M_{\rm vir}$ refers to the total mass enclosed within $r_{\rm vir}$, the radius within which the mean matter density equals $200$ times the critical density of the universe. As described in \citet{fang20}, we determine the concentration $c$ and then the scale radius $r_{\rm s}\equiv r_{\rm vir}/c $ according to the correlation between $c$ and $M_{\rm vir}$ derived from cosmological simulations \citep{duffy08}. 

In our default models, we adopt a physically-motivated corona density profile \citep{fang20}: 
\begin{eqnarray}
\rho(r)=\frac{\rho_{0}}{(r+r_{1})(r+r_{\text{2}})^{2}}\text{~,}  
\end{eqnarray}
\noindent
where $\rho_{0}$ is a constant normalization, $r_{1}$ represents an inner core whose value is chosen to be $r_{1}=3r_{\rm s}/4$ as suggested by cosmological simulations \citep{maller04}, and $r_{2}$ represents the impact of Galactic feedback processes on the halo gas distribution \citep{mathews17,fang20}. When $r_{2}=r_{\rm s}$, this profile reduces to a cored NFW distribution, representing the case without any impact of feedback processes. AGN and stellar feedback processes are expected to deposit energy and momentum into the gaseous halo, heating the gas and pushing the halo gas outward, leading to $r_{2}>r_{\rm s}$. We consider density profiles with a large range of $r_{2}$ ($100 \lesssim r_{2} \lesssim 300$ kpc), which are roughly consistent with the $\beta$ model ($\rho \propto r^{-1.5}$) suggested by observations \citep{miller15,bregman18} at Galactocentric distances of a few tens to $\sim200$ kpc. Our density distribution is flat at $r\ll r_{1}$, and scales roughly as $\rho \propto r^{-1}$ at $r_{1} \ll r \ll r_{2}$. At sufficiently large radii $r \gg  r_{2}$, it approaches to the reduced NFW distribution: $\rho(r) \propto r^{-3}$, guaranteeing that distant regions are not substantially affected by feedback processes.

We determine the normalization of the corona density profile with the electron number density $n_{\rm e}=9.3\times 10^{-5}$ cm$^{-3}$ at $r=59$ kpc, which is the average density from two recent estimates based on the ram-pressure stripping models of MW satellites:  $n_{\rm e}=(6.8$-$18.8) \times 10^{-5}$ cm$^{-3}$ at $r=70 \pm 20$ kpc from \citet{gatto13} and $n_{\rm e}=(3.4$-$8.0) \times 10^{-5}$ cm$^{-3}$ at $r=48.2\pm 2.5$ kpc from \citet{salem15}. Here we have converted the estimated total number densities in these two references to $n_{\rm e}$, which is related with $\rho$ via $\rho=\mu_{\rm e}n_{\rm e} m_{\mu}$, where $\mu_{\rm e}=1.17$ is the mean molecular weight per electron \citep{guo18,zhang20}. We note that the density normalization (i.e., the total corona mass) has no impact on the derived gas temperature profile and thus the constraint on $M_{\rm vir}$, as Equation (\ref{hydro1}) contains the density slope, but not its normalization.

  \begin{figure*}
   \begin{center}
\includegraphics[height=0.25\textwidth]{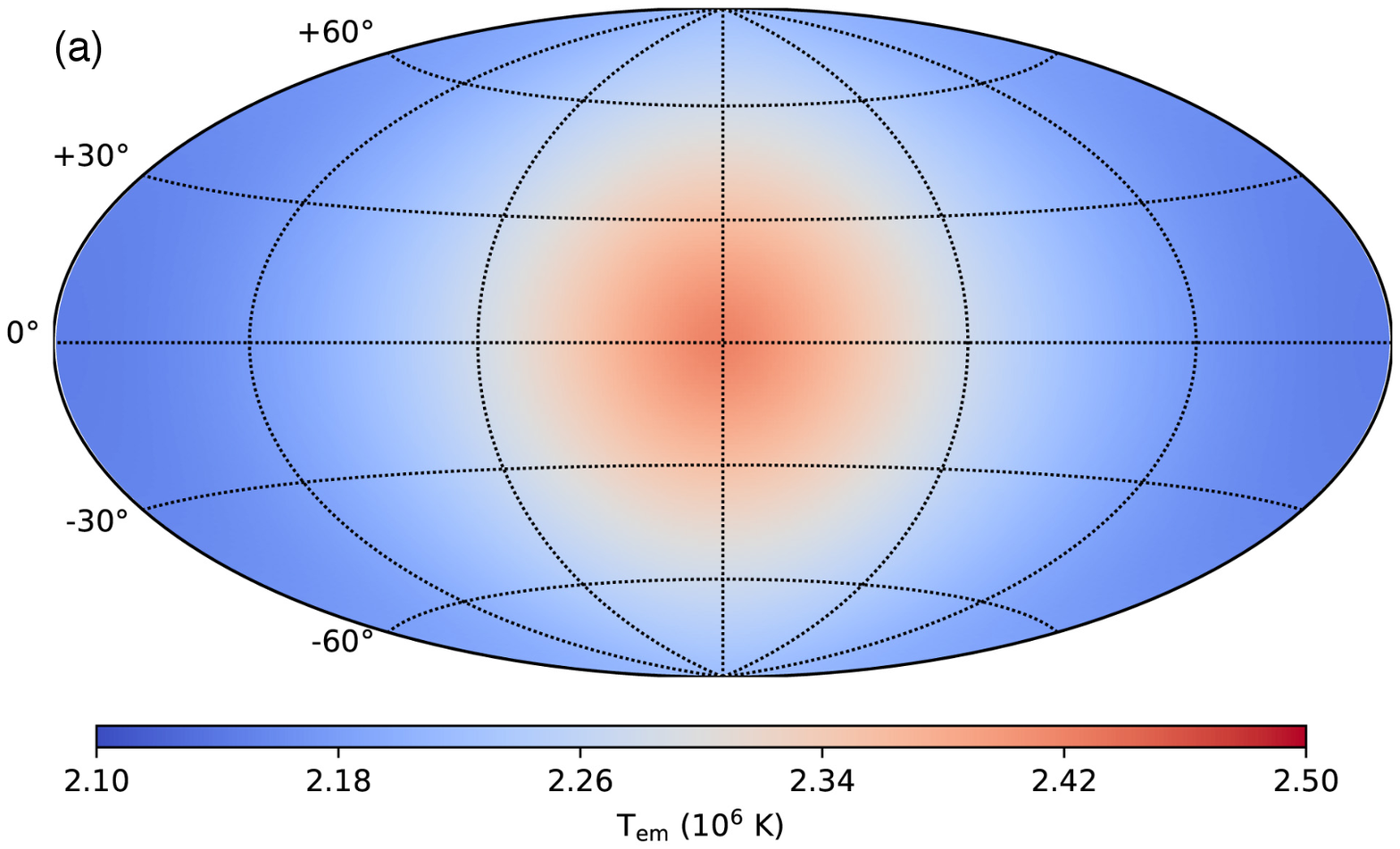} ~~~~~~~~~~~~~~
\includegraphics[height=0.25\textwidth]{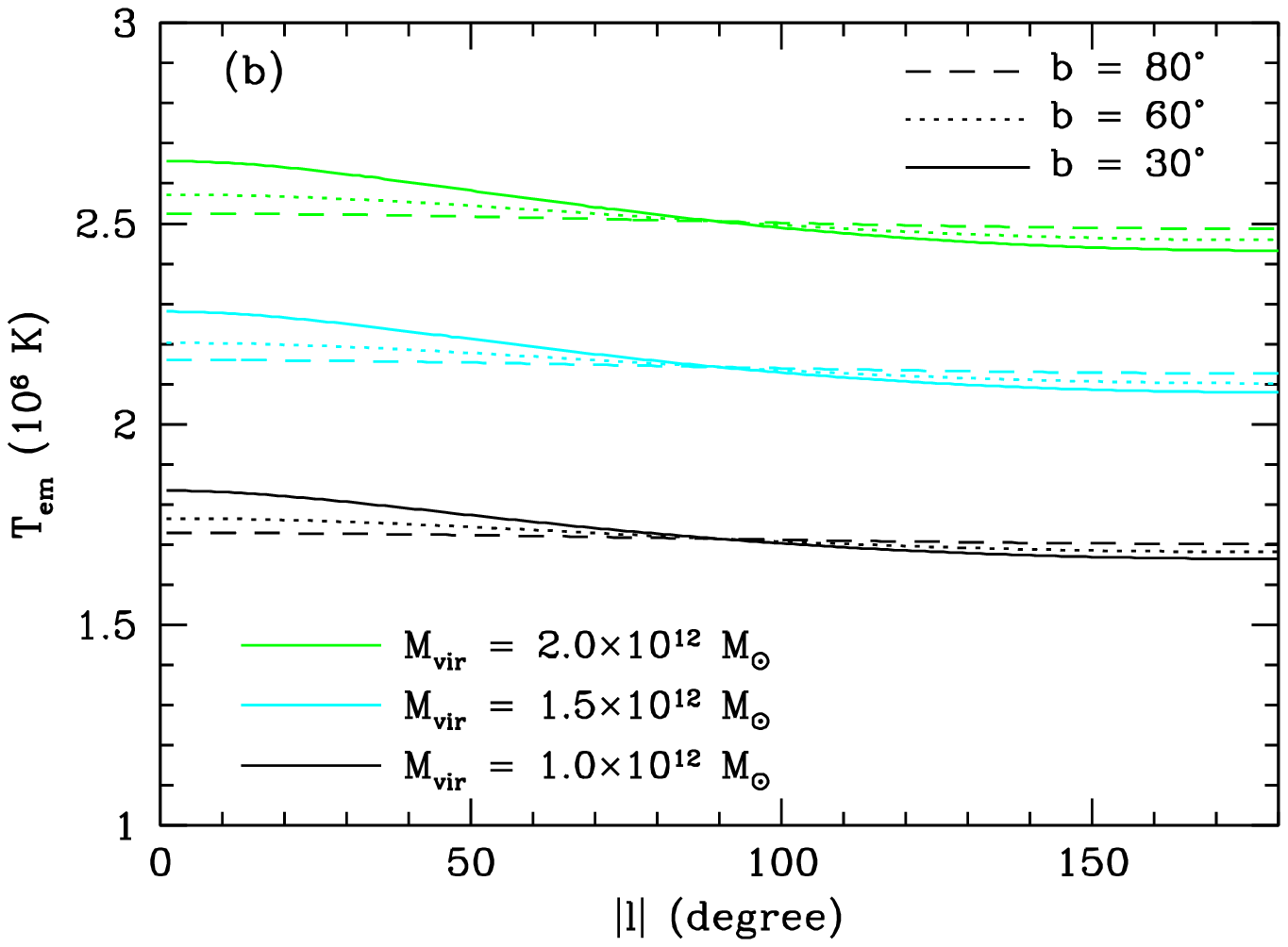} 
  \end{center}
\vspace{-0.5cm}
\caption{Line-of-sight averaged gas temperature distribution $T_{\rm em}$ in Galactic coordinates. (a) $T_{\rm em}$ in our baseline model with $M_{\rm vir}=1.60\times 10^{12} M_{\sun}$, $f_{\rm nt}=0$, and $r_{2}=200$ kpc which results in a characteristic value of $T_{\rm em}$ along $l=90^{\circ}$ equal to $T_{\rm obs}=2.22\times 10^{6}$ K in \citetalias{henley13}.
$T_{\rm em}$ is quite uniform along different sight lines, increasing slightly toward the GC. $T_{\rm em}$ toward the GC region is expected to be further affected by Galactic feedback processes, such as the Fermi bubbles \citep{bland03,su2010,zhang20}. (b) Dependence of $T_{\rm em}$ on $M_{\rm vir}$. Here $T_{\rm em}$ is shown as a function of Galactic longitude at three Galactic latitudes $b=30^{\circ}$ (solid), $60^{\circ}$ (dotted), and $80^{\circ}$ (dashed). The top green, middle cyan, and bottom black lines refer to models with $M_{\rm vir}=2\times 10^{12} M_{\sun}$, $1.5\times 10^{12} M_{\sun}$, $10^{12} M_{\sun}$, respectively. Default values of $f_{\rm nt}=0$ and $r_{2}=200$ kpc are adopted in these models.}
 \label{plot2}
 \end{figure*}

\section{Results}

\subsection{The Corona Temperature Distribution}

We first consider models with the frequently-adopted MW mass $M_{\rm vir} = 10^{12} M_{\sun}$ (\citetalias{bland16}; \citealt{wang20}), which leads to $r_{\rm vir}=207$ kpc, $c=6.36$, and $r_{\rm s}=32.5$ kpc. Fig. 1(a) shows radial profiles of electron number density and temperature in five representative models with varying values of $r_{2}$ from $100$ to $300$ kpc and $f_{\rm nt}$ from $0$ to $0.2$. The dotted, short-dashed, and long-dashed lines demonstrate that as $r_{2}$ increases, the hot gas is distributed more extendedly and the density slope $-d\text{~ln}\rho/d \text{~ln}r$ drops. According to Equation (1), the temperature slope $-d\text{~ln}T/d \text{~ln}r$ increases, leading to an increase in the gas temperature in the inner region. Similarly, an increase in $f_{\rm nt}$ leads to a decrease in $T$ in the inner region. The solid line shows a model with a constant non-thermal pressure fraction $f_{\rm nt}=0.2$, which results in substantially lower gas temperatures compared to the corresponding hydrostatic model with $f_{\rm nt}=0$ and the same density profile. The dot-dashed line refers to a model with $f_{\rm nt}=0.2$ at $r\leq 50$ kpc and $0$ at larger radii, which has similar gas temperatures in the inner region as the model with a radially constant value of $f_{\rm nt}=0.2$. Remarkably, in all theses five models, the gas temperatures in the halo are typically lower than the observed value of $T_{\rm obs}\sim 0.2$ keV (\citealt{henley13}, hereafter HS13; \citealt{yoshino09}). 

We explored the parameter space of our default model and found that the temperature distribution is strongly affected by $M_{\rm vir}$, as illustrated in Fig. 1(b). As implied in Equation (1), $M_{\rm vir}$ determines the gravitational potential well of the halo and thus significantly affects the equilibrium gas temperature distribution, while its impact on our model density profile is negligible. As $M_{\rm vir}$ increases from $10^{12} M_{\sun}$ to $2\times 10^{12} M_{\sun}$, the central gas temperature roughly increases from $2\times 10^{6}$ K to $3\times 10^{6}$ K. The corona density distribution, characterized by $r_{2}$, plays a minor role in determining the derived equilibrium temperature distribution, as seen in Fig. 1(a).  
 
We also applied our calculations to the $\beta$ model of the corona density distribution $\rho(r)=\rho_{0}(1+(r/r_{\rm c})^{2})^{-3\beta/2}$, where $\rho_{0}$ is the core density, $r_{\rm c}$ is the core radius, and $-3\beta$ is the slope of the profile at large radii. Following recent X-ray observations \citep{miller15,bregman18}, we adopt $\beta=0.5$. Several representative density and temperature profiles of this model are shown in Fig. 1(c). At $r\gtrsim 1$ kpc, the $\beta$ model with $r_{\rm c}=0.1$ kpc is essentially the same as the power-law profile ($\rho \propto r^{-1.5}$) frequently used in X-ray observations \citep{miller15,bregman18}. This model leads to an equilibrium temperature profile decreasing inwards in the inner region ($r\lesssim 40$ kpc) due to high density slopes there. As $r_{\rm c}$ increases, the inner density slope decreases and $T$ in the inner region increases. In general, the $\beta$ model is not isothermal as assumed in many observations \citep{bregman18}.

\subsection{Constraint on the Milky Way Mass}

A comparison between the predicted halo gas temperature with the observed value may thus be used to constrain the MW mass $M_{\rm vir}$. To this end, we adopt the Astrophysical Plasma Emission Code (APEC; \citealt{smith01,foster12}) to calculate the average gas temperatures $T_{\rm em}$ along individual sight lines weighted by the $0.5-2.0$ keV X-ray emission: 
\begin{eqnarray}
T_{\rm em}(l,b)=\frac{\int_{\rm los}  n_{\rm e} n_{\rm H}T \epsilon(T,Z)dR}{ \int_{\rm los} n_{\rm e}n_{\rm H}\epsilon(T,Z)dR} \text{,} 
\end{eqnarray}
\noindent
where $\epsilon(T,Z)$ is the $0.5-2.0$ keV X-ray emissivity, and $l$ and $b$ refer to the Galactic longitude and latitude, respectively. The distance $R$ of each gas element to the Earth is related to its Galactocentric distance $r$ via $r^{2}=R^{2}+R_{\sun}^{2}-2R_{\sun}R \text{~cos~} l \text{~cos~} b$, where $R_{\sun}=8.5$ kpc is the distance between the Earth and the GC. Along each line of sight, the integration is done to a distance of 240 kpc from the Earth. The hot gas is assumed to be optically thin and under collisional ionization equilibrium, and as in \citetalias{henley13}, we adopt the solar metallicity $Z=Z_{\odot}$.

Although $T(r)$ varies substantially along the radial direction in our models (see Fig. 1), the line-of-sight averaged temperature $T_{\rm em}$ varies very little across different sight lines (typically $<10\%$ at$|b|>30^{\circ}$), as clearly illustrated in Fig. 2 and consistent with the observed fairly uniform gas temperature $T_{\rm obs}\sim 0.2$ keV in both Suzaku \citep{yoshino09} and XMM-Newton observations \citepalias{henley13}. This is merely due to the fact that $T(r)$ is spherically-symmetric and $R_{\sun}$ is very small compared to the halo size. Fig. 2(b) shows the variations of $T_{\rm em}$ as a function of Galactic longitude and latitude for three models with different MW masses. While $T_{\rm em}$ varies very little with Galactic latitude and longitude, it increases significantly with $M_{\rm vir}$. Our calculations thus indicate that the observed fairly uniform gas temperature toward different sight lines does not preclude substantial radial variations in the corona temperature distribution. 
 
\begin{figure}
    \begin{center}
\includegraphics[width=0.45\textwidth]{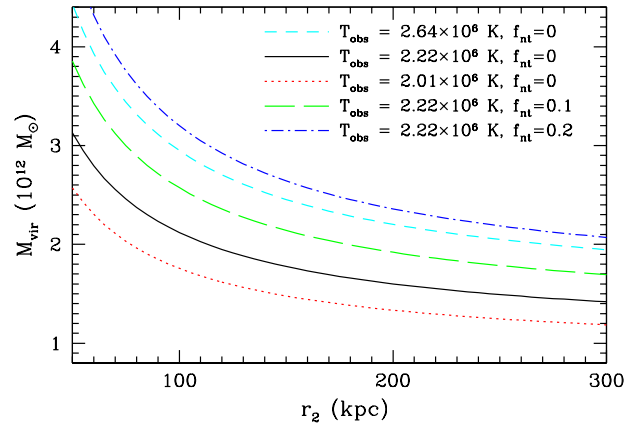} 
  \end{center}
\vspace{-0.5cm}
\caption{$M_{\rm vir}$ constrained by the condition $T_{\rm em}(l=90^{\circ})=T_{\rm obs}$. $M_{\rm vir}$ is a function of $r_{2}$ and $f_{\rm nt}$. The solid line represents the case with ${T}_{\rm obs}=2.22\times 10^{6}$ K, the median temperature measured by {\it XMM-Newton} observations \citepalias{henley13}. The short-dashed and dotted lines refer to the measured upper-quartile and lower-quartile temperatures: ${T}_{\rm obs}=2.64\times 10^{6}$ K, $2.01\times 10^{6}$ K, respectively.  }
 \label{plot3}
 \end{figure}

To constrain $M_{\rm vir}$, we use the predicted value of $T_{\rm em}$ along $l=90^{\circ}$, which is independent of $b$ and is roughly the mean value of $T_{\rm em}$ along all the sightlines. We first consider models with $f_{\rm nt}=0$ and take $M_{\rm vir}$ and $r_{2}$ as the two main model parameters. For a given value of $r_{2}$ within $100 \text{-} 300$ kpc, we determine $M_{\rm vir}$ so that $T_{\rm em}(l=90^{\circ})$ equals $T_{\rm obs}=(2.01$ - $2.64)\times 10^{6}$ K measured by \citetalias{henley13}. Therefore we constrain the MW mass to be $M_{\rm vir}=(1.19$ - $2.95) \times 10^{12} M_{\sun}$. Fig. \ref{plot3} further shows that the uncertainties in $M_{\rm vir}$ mainly come from those in $T_{\rm obs}$, while the corona density profile plays a minor role. As $r_{2}$ increases, the density slope drops and the equilibrium corona temperature increases, resulting in a decrease in the derived value of $M_{\rm vir}$. Non-thermal pressure support leads to higher values of $M_{\rm vir}$ (see Fig. \ref{plot3}), and for $T_{\rm obs}=2.22\times 10^{6}$ K and $r_{2}=200$ kpc, $M_{\rm vir}$ increases by $ 20\%$ and $47\%$ if $f_{\rm nt}$ increases from $0$ to $0.1$ and $0.2$, respectively.

Considering a baseline model with $T_{\rm obs}=2.22\times 10^{6}$ K (the median temperature measured by \citetalias{henley13}), $f_{\rm nt}=0$, and $r_{2}=200$ kpc, we have $M_{\rm vir}=1.60\times 10^{12} M_{\sun}$, and subsequently, $r_{\rm vir}=242$ kpc, $c=6.07$, $r_{\rm s}=39.8$ kpc, and a local DM density at the solar position of $0.22$ GeV cm$^{-3}$. The total hot gas mass within $r_{\rm vir}$ is $3.8\times 10^{10} M_{\sun}$. Taking the cold baryonic mass (stars and cold gas) of the MW to be $M_{\rm ctot} \approx 6 \times 10^{10} M_{\sun}$ (\citetalias{bland16}; \citealt{McMillan2017}), the total baryonic mass is $9.8\times 10^{10} M_{\sun}$. However, according to the cosmic baryon fraction $f_{\rm b}=0.157$ \citep{planck16}, the MW's baryonic allotment should be $M_{\rm b}=f_{\rm b}M_{\rm vir}=2.51\times 10^{11} M_{\sun}$. Therefore, the baryonic mass missing within $r_{\rm vir}$ is $1.53\times 10^{11} M_{\sun}$ (about $61\%$), potentially residing beyond $r_{\rm vir}$ or in a cool phase in the halo.

\section{Discussions}

Our method relies on the assumption that the diffuse corona extends to the outer regions of the MW halo, which is predicted in galaxy formation simulations (e.g., \citealt{crain10}; \citealt{sok16}). While this is consistent with the ram-pressure stripping and gas cloud confinement arguments \citep{fang13}, it is still unclear if the halo X-ray emission is mainly contributed by a spherical corona (\citetalias{henley13}; \citealt{miller15}) or a disk-like gas distribution with a scale height of a few kiloparsecs \citep{yao09,nakashima18}. Our calculations support the former picture, while a significant contribution of the latter to the halo X-ray emission can not be ruled out. In our baseline model where the corona density profile is normalized by the recent density estimates from the ram-pressure stripping models, the predicted $0.5-2.0$ keV X-ray surface brightness typically ranges from $1.4\times 10^{-12}$ erg cm$^{-2}$ s$^{-1}$ deg$^{-2}$ along the $(l,b)=(180^{\circ},30^{\circ})$ sightline to $1.9\times 10^{-12}$ erg cm$^{-2}$ s$^{-1}$ deg$^{-2}$ along the $(l,b)=(180^{\circ},90^{\circ})$ sightline, consistent with the typical values of $(1.1$ - $2.3)\times 10^{-12}$ erg cm$^{-2}$ s$^{-1}$ deg$^{-2}$ measured by \citetalias{henley13}. Similarly, the predicted emission measures of $(2.1$ - $2.7)\times 10^{-3}$ cm$^{-6}$ pc are also consistent with the typical values of $(1.4$ - $3.0)\times 10^{-3}$ cm$^{-6}$ pc observed by \citetalias{henley13}. The predicted $0.5-2.0$ keV X-ray luminosities within $r\leq 50$ and $200$ kpc are $3.35\times 10^{39}$ erg s$^{-1}$ and $6.04\times 10^{39}$ erg s$^{-1}$, respectively.

We also assume that the corona gas is in a dynamical equilibrium state described by Equation (\ref{hydro1}), which incorporates potential non-thermal pressure support from radial and rotating bulk motions, turbulent motions, cosmic rays, and magnetic fields \citep{hk16,oppenheimer18}. In galaxy clusters, hydrodynamic simulations suggest that non-thermal pressure support typically causes an underestimate of the real cluster mass by about $10-20\%$ \citep{nelson14,biffi16}, but recent X-ray observations \citep{eckert19} imply a substantially lower non-thermal pressure fraction $f_{\rm nt}\sim 6-10\%$. A lower value of $M_{\rm vir}$ may be possible if the corona gas in most regions within $50$ kpc is outflowing acceleratingly, which tends to counteract the impact of non-thermal pressure support. However, the star formation activity in the GC has been very quiescent during most times of the past 8 Gyr \citep{Nogueras2019} and the X-ray measurements of $T_{\rm obs}\sim 0.2$ keV have avoided the sightlines toward the inner Galaxy and other regions with strong stellar feedback. 

The adopted value of gas metallicity has nearly no impact on the constrained MW mass, as the X-ray emissivity $\epsilon(T,Z)$ appears in both the denominator and numerator in the right hand side of Equation (3). For the baseline model, lower values of $Z=0.5Z_{\odot}$ and $0.3Z_{\odot}$ lead to a negligible decrease in $M_{\rm vir}$ by $0.05\%$ and $0.12\%$, respectively. $M_{\rm vir}$ is also independent of the adopted outer gas temperature $T_{\rm out}$, which mainly affects temperature at large radii. $T_{\rm em}$ is mainly determined by the inner region $R_{\sun}<r< 50$ kpc, which contributes to $\sim 95\%$ of the $0.5-2.0$ keV X-ray surface brightness along a representative sight line toward $l=90^{\circ}$ in our baseline model. Within this region, Equation (1) leads to $P(r)=P(r_{\rm out})+\int_{r}^{r_{\rm out}}(1-f_{\rm nt})\rho \frac{GM(r)}{r^{2}}dr \approx \int_{r}^{r_{\rm out}}(1-f_{\rm nt})\rho \frac{GM(r)}{r^{2}}dr$ as $P(r_{\rm out})$ is typically lower than $P(r)$ by two orders of magnitude due to the fast decreasing of $\rho$ at large radii.

Our results are quite robust to the adopted corona density profile. We applied our calculations to the $\beta$ model, taking $f_{\rm nt}=0$ and $T_{\rm em}(l=90^{\circ})=2.22\times 10^{6}$ K. For models with a core radius of $r_{\rm c}\sim 20-30$ kpc (close to $r_{1}=3r_{\rm s}/4$ in our default models), the inner density profile is flat, and the derived value of $M_{\rm vir}$ decreases from $M_{\rm vir}=2.08\times 10^{12}M_{\sun}$ if $r_{\rm c}=20$ kpc to $M_{\rm vir}=1.42\times 10^{12}M_{\sun}$ if $r_{\rm c}=30$ kpc, consistent with our previous results. In contrast, a cuspy density profile with $r_{\rm c}=0.1$ kpc would lead to low inner gas temperatures (Fig. 1c) and therefore a high value of $M_{\rm vir}=7.38\times 10^{12}M_{\sun}$, inconsistent with current measurements of $M_{\rm vir}=(0.5 \text{-} 2)\times 10^{12}M_{\sun}$ (\citetalias{bland16}; \citealt{wang20}).

\begin{figure}
  \begin{center}
\includegraphics[width=0.45\textwidth]{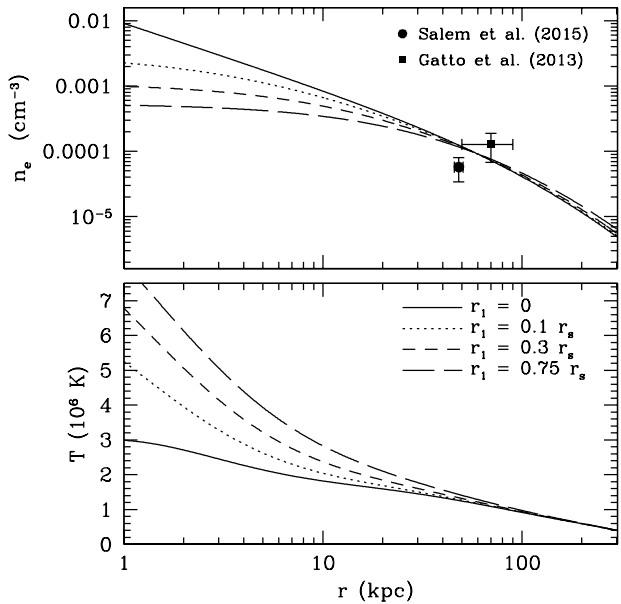}
  \end{center}
\vspace{-0.5cm}
\caption{Radial profiles of gas density and temperature in four models where $M(r)$ is contributed by an NFW DM profile with $M_{\rm vir,dm}=10^{12}M_{\sun}$ and a central baryonic matter distribution with $M_{\rm ctot}= 6 \times 10^{10} M_{\sun}$. $f_{\rm nt}=0$ and $r_{2}=200$ kpc are adopted in these models. As $r_{1}$ increases, the gas density distribution becomes more spatially extended, and the temperature in the inner region increases.}  
 \label{plot4}
\end{figure}
 
A cuspy corona density profile is possible if the total matter distribution $M(r)$ is more centrally-peaked than the adopted NFW profile. In the central region, cold baryons contribute significantly to $M(r)$ and baryonic physics may also cause contraction or expansion of the DM halo \citep{blumenthal86,navarro96,marinacci14,dutton16}. Here we consider an additional case where $M(r)$ is contributed by an NFW DM distribution with the virial mass $M_{\rm vir,dm}$ and a central cold baryonic matter distribution with $M_{\rm ctot}= 6 \times 10^{10} M_{\sun}$. The latter is approximated with a Hernquist profile $M_{\rm cold}(r)=M_{\rm ctot}r^{2}/(r+a)^{2}$ \citep{hernquist90}, where $a$ is chosen to be $1.5$ kpc so that the resulting gravitational acceleration fits reasonably well with that in the more realistic model in \citet{McMillan2017} and \citet{zhang20}. To offset stronger gravity in this case, higher pressure gradients are required in the inner region. For cored corona density profiles with $r_{1}\gtrsim 0.1r_{\rm s}$, this leads to cuspy temperature profiles with inner gas temperatures higher than $4\times 10^{6}$ K, as clearly shown in Fig. 4. However, X-ray observations indicate that the corona temperature in the inner region is about $0.3$ keV $\approx 3.5 \times 10^{6}$ K \citep{Kataoka2013,Kataoka2015}. Considering ongoing feedback heating processes there \citep{bland03,su2010,guo12,zhang20}, the equilibrium temperature may be even lower, which is viable if the inner corona density profile is cuspy with $r_{1}=0$ ($\rho \propto r^{-1}$), as shown in Fig. 4. In this case, the constrained DM virial mass by $T_{\rm obs}=(2.01$ - $2.64)\times 10^{6}$ K and $r_{2} =100$ - $300$ kpc is $M_{\rm vir,dm}=(1.34$ - $5.44) \times 10^{12} M_{\sun}$, and in the baseline model, we derive $M_{\rm vir,dm}=2.13 \times 10^{12} M_{\sun}$.

The uncertainty in our constraint on $M_{\rm vir}$ mainly comes from the corona temperature measurement. Recently, the Suzaku X-ray observations measured a substantially higher value for $T_{\rm obs}=3.0\times 10^{6}$ K \citep{nakashima18}, which corresponds to $M_{\rm vir}=2.8\times 10^{12} M_{\sun}$ if $f_{\rm nt}=0$, and $r_{2}=200$ kpc. This value is substantially higher than $M_{\rm vir}=1.60\times 10^{12} M_{\sun}$ in our baseline model constrained by $T_{\rm obs}=2.22\times 10^{6}$ K. Recent X-ray observations also suggest that multiple temperature components may exist along some sightlines \citep{das19}, and the hot components with $T\sim 10^{7}$ K may be associated with local stellar or black hole feedback processes such as the Fermi bubbles, which if true, does not substantially affect our results.

\section{Conclusions and Outlook}

We propose a novel and independent method to constrain the MW mass based on its corona temperature observations. We identify two classes of equilibrium models consistent with current observations: (1) For an NFW total matter distribution, the corona density profile should be cored, and the MW mass is constrained to be $M_{\rm vir}=(1.19$ - $2.95) \times 10^{12} M_{\sun}$; (2) For a total matter distribution contributed by an NFW DM distribution and a central baryon distribution, the corona density profile should be cuspy, and $M_{\rm vir,dm}=(1.34$ - $5.44) \times 10^{12} M_{\sun}$. Both constraints overlap with the estimates of $M_{\rm vir}=(0.5$ - $2)\times 10^{12}M_{\sun}$ in the literature (\citealt{xue08}; \citetalias{bland16}; \citealt{lizz17}; \citealt{wang20}), and lie on the high mass side $M_{\rm vir}>10^{12}M_{\sun}$. Non-thermal pressure support, which likely exists in the corona, would lead to even higher values of $M_{\rm vir}$, and for the former case, $M_{\rm vir}=(1.42$ - $3.59)\times 10^{12} M_{\sun}$ if $f_{\rm nt}=0.1$. 

Our estimate of $M_{\rm vir}$ implies that the Magellanic Clouds and the Leo I dwarf spheroidal are bound to the MW \citep{boylan13,cautun14}, a large fraction of the baryons are missing in the MW, and the ``too-big-to-fail" problem may pose a challenge to the cold DM theory \citep{boylan12}. The uncertainty in our constraint on $M_{\rm vir}$ comes from the uncertainties in $T_{\rm obs}$ and the corona density profile. Zoom-in cosmological simulations of MW-like galaxies are expected to improve our understanding of the corona dynamical state and its temperature and density distributions, helping further constrain $M_{\rm vir}$. The SRG/eROSITA telescope is currently taking a sensitive full-sky X-ray survey with X-ray spectra taken automatically along all the sight lines, which may statistically improve the measurement of $T_{\rm obs}$ and its variations with Galactic latitude and longitude, increasing the accuracy of the X-ray constraint on $M_{\rm vir}$.

We thank an anonymous referee for very insightful comments. This work was supported by the National Natural Science Foundation of China (No. 11873072 and 11633006), the Natural Science Foundation of Shanghai (No. 18ZR1447100), and Chinese Academy of Sciences through the Key Research Program of Frontier Sciences (No. QYZDB-SSW-SYS033 and QYZDJ-SSW-SYS008).

\bibliography{ms}

\begin{thebibliography}{}
\expandafter\ifx\csname natexlab\endcsname\relax\def\natexlab#1{#1}\fi
\providecommand{\url}[1]{\href{#1}{#1}}

\bibitem[{{Allen} {et~al.}(2011){Allen}, {Evrard}, \& {Mantz}}]{allen11}
{Allen}, S.~W., {Evrard}, A.~E., \& {Mantz}, A.~B. 2011, \araa, 49, 409

\bibitem[{{Biffi} {et~al.}(2016){Biffi}, {Borgani}, {Murante}, {Rasia},
  {Planelles}, {Granato}, {Ragone-Figueroa}, {Beck}, {Gaspari}, \&
  {Dolag}}]{biffi16}
{Biffi}, V., {Borgani}, S., {Murante}, G., {et~al.} 2016, \apj, 827, 112

\bibitem[{{Binney} \& {Tremaine}(2008)}]{binney08}
{Binney}, J., \& {Tremaine}, S. 2008, {Galactic Dynamics: Second Edition}
  (Princeton: Princeton University Press)

\bibitem[{{Bland-Hawthorn} \& {Cohen}(2003)}]{bland03}
{Bland-Hawthorn}, J., \& {Cohen}, M. 2003, \apj, 582, 246

\bibitem[{{Bland-Hawthorn} \& {Gerhard}(2016)}]{bland16}
{Bland-Hawthorn}, J., \& {Gerhard}, O. 2016, \araa, 54, 529

\bibitem[{{Blumenthal} {et~al.}(1986){Blumenthal}, {Faber}, {Flores}, \&
  {Primack}}]{blumenthal86}
{Blumenthal}, G.~R., {Faber}, S.~M., {Flores}, R., \& {Primack}, J.~R. 1986,
  \apj, 301, 27

\bibitem[{{Boylan-Kolchin} {et~al.}(2012){Boylan-Kolchin}, {Bullock}, \&
  {Kaplinghat}}]{boylan12}
{Boylan-Kolchin}, M., {Bullock}, J.~S., \& {Kaplinghat}, M. 2012, \mnras, 422,
  1203

\bibitem[{{Boylan-Kolchin} {et~al.}(2013){Boylan-Kolchin}, {Bullock}, {Sohn},
  {Besla}, \& {van der Marel}}]{boylan13}
{Boylan-Kolchin}, M., {Bullock}, J.~S., {Sohn}, S.~T., {Besla}, G., \& {van der
  Marel}, R.~P. 2013, \apj, 768, 140

\bibitem[{{Bregman} {et~al.}(2018){Bregman}, {Anderson}, {Miller},
  {Hodges-Kluck}, {Dai}, {Li}, {Li}, \& {Qu}}]{bregman18}
{Bregman}, J.~N., {Anderson}, M.~E., {Miller}, M.~J., {et~al.} 2018, \apj, 862,
  3

\bibitem[{{Cautun} {et~al.}(2014){Cautun}, {Frenk}, {van de Weygaert},
  {Hellwing}, \& {Jones}}]{cautun14}
{Cautun}, M., {Frenk}, C.~S., {van de Weygaert}, R., {Hellwing}, W.~A., \&
  {Jones}, B. J.~T. 2014, \mnras, 445, 2049

\bibitem[{{Crain} {et~al.}(2010){Crain}, {McCarthy}, {Frenk}, {Theuns}, \&
  {Schaye}}]{crain10}
{Crain}, R.~A., {McCarthy}, I.~G., {Frenk}, C.~S., {Theuns}, T., \& {Schaye},
  J. 2010, \mnras, 407, 1403

\bibitem[{{Das} {et~al.}(2019){Das}, {Mathur}, {Gupta}, {Nicastro}, \&
  {Krongold}}]{das19}
{Das}, S., {Mathur}, S., {Gupta}, A., {Nicastro}, F., \& {Krongold}, Y. 2019,
  \apj, 887, 257

\bibitem[{{Duffy} {et~al.}(2008){Duffy}, {Schaye}, {Kay}, \& {Dalla
  Vecchia}}]{duffy08}
{Duffy}, A.~R., {Schaye}, J., {Kay}, S.~T., \& {Dalla Vecchia}, C. 2008,
  \mnras, 390, L64

\bibitem[{{Dutton} {et~al.}(2016){Dutton}, {Macci{\`o}}, {Dekel}, {Wang},
  {Stinson}, {Obreja}, {Di Cintio}, {Brook}, {Buck}, \& {Kang}}]{dutton16}
{Dutton}, A.~A., {Macci{\`o}}, A.~V., {Dekel}, A., {et~al.} 2016, \mnras, 461,
  2658

\bibitem[{{Eckert} {et~al.}(2019){Eckert}, {Ghirardini}, {Ettori}, {Rasia},
  {Biffi}, {Pointecouteau}, {Rossetti}, {Molendi}, {Vazza}, {Gastaldello},
  {Gaspari}, {De Grandi}, {Ghizzardi}, {Bourdin}, {Tchernin}, \&
  {Roncarelli}}]{eckert19}
{Eckert}, D., {Ghirardini}, V., {Ettori}, S., {et~al.} 2019, \aap, 621, A40

\bibitem[{{Fang} {et~al.}(2013){Fang}, {Bullock}, \& {Boylan-Kolchin}}]{fang13}
{Fang}, T., {Bullock}, J., \& {Boylan-Kolchin}, M. 2013, \apj, 762, 20

\bibitem[{{Fang} {et~al.}(2020){Fang}, {Guo}, \& {Yuan}}]{fang20}
{Fang}, X.-E., {Guo}, F., \& {Yuan}, Y.-F. 2020, \apj, 894, 1

\bibitem[{{Foster} {et~al.}(2012){Foster}, {Ji}, {Smith}, \&
  {Brickhouse}}]{foster12}
{Foster}, A.~R., {Ji}, L., {Smith}, R.~K., \& {Brickhouse}, N.~S. 2012, \apj,
  756, 128

\bibitem[{{Gatto} {et~al.}(2013){Gatto}, {Fraternali}, {Read}, {Marinacci},
  {Lux}, \& {Walch}}]{gatto13}
{Gatto}, A., {Fraternali}, F., {Read}, J.~I., {et~al.} 2013, \mnras, 433, 2749

\bibitem[{{Ghirardini} {et~al.}(2019){Ghirardini}, {Eckert}, {Ettori},
  {Pointecouteau}, {Molendi}, {Gaspari}, {Rossetti}, {De Grandi}, {Roncarelli},
  {Bourdin}, {Mazzotta}, {Rasia}, \& {Vazza}}]{ghirardini19}
{Ghirardini}, V., {Eckert}, D., {Ettori}, S., {et~al.} 2019, \aap, 621, A41

\bibitem[{{Guo} {et~al.}(2018){Guo}, {Duan}, \& {Yuan}}]{guo18}
{Guo}, F., {Duan}, X., \& {Yuan}, Y.-F. 2018, \mnras, 473, 1332

\bibitem[{{Guo} \& {Mathews}(2012)}]{guo12}
{Guo}, F., \& {Mathews}, W.~G. 2012, \apj, 756, 181

\bibitem[{{Henley} \& {Shelton}(2013)}]{henley13}
{Henley}, D.~B., \& {Shelton}, R.~L. 2013, \apj, 773, 92

\bibitem[{{Hernquist}(1990)}]{hernquist90}
{Hernquist}, L. 1990, \apj, 356, 359

\bibitem[{{Hodges-Kluck} {et~al.}(2016){Hodges-Kluck}, {Miller}, \&
  {Bregman}}]{hk16}
{Hodges-Kluck}, E.~J., {Miller}, M.~J., \& {Bregman}, J.~N. 2016, \apj, 822, 21

\bibitem[{{Kataoka} {et~al.}(2015){Kataoka}, {Tahara}, {Totani}, {Sofue},
  {Inoue}, {Nakashima}, \& {Cheung}}]{Kataoka2015}
{Kataoka}, J., {Tahara}, M., {Totani}, T., {et~al.} 2015, \apj, 807, 77

\bibitem[{{Kataoka} {et~al.}(2013){Kataoka}, {Tahara}, {Totani}, {Sofue},
  {Stawarz}, {Takahashi}, {Takeuchi}, {Tsunemi}, {Kimura}, {Takei}, {Cheung},
  {Inoue}, \& {Nakamori}}]{Kataoka2013}
---. 2013, \apj, 779, 57

\bibitem[{{Kravtsov} \& {Borgani}(2012)}]{kravtsov12}
{Kravtsov}, A.~V., \& {Borgani}, S. 2012, \araa, 50, 353

\bibitem[{{Li} {et~al.}(2017){Li}, {Jing}, {Qian}, {Yuan}, \& {Zhao}}]{lizz17}
{Li}, Z.-Z., {Jing}, Y.~P., {Qian}, Y.-Z., {Yuan}, Z., \& {Zhao}, D.-H. 2017,
  \apj, 850, 116

\bibitem[{{Maller} \& {Bullock}(2004)}]{maller04}
{Maller}, A.~H., \& {Bullock}, J.~S. 2004, \mnras, 355, 694

\bibitem[{{Marinacci} {et~al.}(2014){Marinacci}, {Pakmor}, \&
  {Springel}}]{marinacci14}
{Marinacci}, F., {Pakmor}, R., \& {Springel}, V. 2014, \mnras, 437, 1750

\bibitem[{{Mathews} \& {Prochaska}(2017)}]{mathews17}
{Mathews}, W.~G., \& {Prochaska}, J.~X. 2017, \apjl, 846, L24

\bibitem[{{McMillan}(2017)}]{McMillan2017}
{McMillan}, P.~J. 2017, \mnras, 465, 76

\bibitem[{{Miller} \& {Bregman}(2015)}]{miller15}
{Miller}, M.~J., \& {Bregman}, J.~N. 2015, \apj, 800, 14

\bibitem[{{Nakashima} {et~al.}(2018){Nakashima}, {Inoue}, {Yamasaki}, {Sofue},
  {Kataoka}, \& {Sakai}}]{nakashima18}
{Nakashima}, S., {Inoue}, Y., {Yamasaki}, N., {et~al.} 2018, \apj, 862, 34

\bibitem[{{Navarro} {et~al.}(1996{\natexlab{a}}){Navarro}, {Eke}, \&
  {Frenk}}]{navarro96}
{Navarro}, J.~F., {Eke}, V.~R., \& {Frenk}, C.~S. 1996{\natexlab{a}}, \mnras,
  283, L72

\bibitem[{{Navarro} {et~al.}(1996{\natexlab{b}}){Navarro}, {Frenk}, \&
  {White}}]{navarro1996}
{Navarro}, J.~F., {Frenk}, C.~S., \& {White}, S.~D.~M. 1996{\natexlab{b}},
  \apj, 462, 563

\bibitem[{{Navarro} {et~al.}(1997){Navarro}, {Frenk}, \& {White}}]{navarro97}
{Navarro}, J.~F., {Frenk}, C.~S., \& {White}, S. D.~M. 1997, \apj, 490, 493

\bibitem[{{Nelson} {et~al.}(2014){Nelson}, {Lau}, \& {Nagai}}]{nelson14}
{Nelson}, K., {Lau}, E.~T., \& {Nagai}, D. 2014, \apj, 792, 25

\bibitem[{{Nogueras-Lara} {et~al.}(2019){Nogueras-Lara}, {Sch{\"o}del},
  {Gallego-Calvente}, {Gallego-Cano}, {Shahzamanian}, {Dong}, {Neumayer},
  {Hilker}, {Najarro}, {Nishiyama}, {Feldmeier-Krause}, {Girard}, \&
  {Cassisi}}]{Nogueras2019}
{Nogueras-Lara}, F., {Sch{\"o}del}, R., {Gallego-Calvente}, A.~T., {et~al.}
  2019, Nature Astronomy, 4

\bibitem[{{Oppenheimer}(2018)}]{oppenheimer18}
{Oppenheimer}, B.~D. 2018, \mnras, 480, 2963

\bibitem[{{Planck Collaboration} {et~al.}(2016){Planck Collaboration}, {Ade},
  {Aghanim}, {Arnaud}, {Ashdown}, {Aumont}, {Baccigalupi}, {Banday},
  {Barreiro}, {Bartlett}, \& et~al.}]{planck16}
{Planck Collaboration}, {Ade}, P.~A.~R., {Aghanim}, N., {et~al.} 2016, \aap,
  594, A13

\bibitem[{{Salem} {et~al.}(2015){Salem}, {Besla}, {Bryan}, {Putman}, {van der
  Marel}, \& {Tonnesen}}]{salem15}
{Salem}, M., {Besla}, G., {Bryan}, G., {et~al.} 2015, \apj, 815, 77

\bibitem[{{Smith} {et~al.}(2001){Smith}, {Brickhouse}, {Liedahl}, \&
  {Raymond}}]{smith01}
{Smith}, R.~K., {Brickhouse}, N.~S., {Liedahl}, D.~A., \& {Raymond}, J.~C.
  2001, \apjl, 556, L91

\bibitem[{{Soko{\l}owska} {et~al.}(2016){Soko{\l}owska}, {Mayer}, {Babul},
  {Madau}, \& {Shen}}]{sok16}
{Soko{\l}owska}, A., {Mayer}, L., {Babul}, A., {Madau}, P., \& {Shen}, S. 2016,
  \apj, 819, 21

\bibitem[{{Su} {et~al.}(2010){Su}, {Slatyer}, \& {Finkbeiner}}]{su2010}
{Su}, M., {Slatyer}, T.~R., \& {Finkbeiner}, D.~P. 2010, \apj, 724, 1044

\bibitem[{{Vikhlinin} {et~al.}(2006){Vikhlinin}, {Kravtsov}, {Forman}, {Jones},
  {Markevitch}, {Murray}, \& {Van Speybroeck}}]{vikhlinin06}
{Vikhlinin}, A., {Kravtsov}, A., {Forman}, W., {et~al.} 2006, \apj, 640, 691

\bibitem[{{Wang} {et~al.}(2020){Wang}, {Han}, {Cautun}, {Li}, \&
  {Ishigaki}}]{wang20}
{Wang}, W., {Han}, J., {Cautun}, M., {Li}, Z., \& {Ishigaki}, M.~N. 2020,
  Science China Physics, Mechanics, and Astronomy, 63, 109801

\bibitem[{{Xue} {et~al.}(2008){Xue}, {Rix}, {Zhao}, {Re Fiorentin}, {Naab},
  {Steinmetz}, {van den Bosch}, {Beers}, {Lee}, {Bell}, {Rockosi}, {Yanny},
  {Newberg}, {Wilhelm}, {Kang}, {Smith}, \& {Schneider}}]{xue08}
{Xue}, X.~X., {Rix}, H.~W., {Zhao}, G., {et~al.} 2008, \apj, 684, 1143

\bibitem[{{Yao} {et~al.}(2009){Yao}, {Wang}, {Hagihara}, {Mitsuda}, {McCammon},
  \& {Yamasaki}}]{yao09}
{Yao}, Y., {Wang}, Q.~D., {Hagihara}, T., {et~al.} 2009, \apj, 690, 143

\bibitem[{{Yoshino} {et~al.}(2009){Yoshino}, {Mitsuda}, {Yamasaki}, {Takei},
  {Hagihara}, {Masui}, {Bauer}, {McCammon}, {Fujimoto}, {Wang}, \&
  {Yao}}]{yoshino09}
{Yoshino}, T., {Mitsuda}, K., {Yamasaki}, N.~Y., {et~al.} 2009, \pasj, 61, 805

\bibitem[{{Zhang} \& {Guo}(2020)}]{zhang20}
{Zhang}, R., \& {Guo}, F. 2020, \apj, 894, 117

\end{thebibliography}

\end{document}